\documentclass[english,twocolumn,reprint]{revtex4}
\usepackage[T1]{fontenc}
\usepackage[latin9]{inputenc}
\usepackage{array}
\usepackage{amstext}
\usepackage{graphicx}

\makeatletter

\newcommand{\lyxmathsym}[1]{\ifmmode\begingroup\def\b@ld{bold}
  \text{\ifx\math@version\b@ld\bfseries\fi#1}\endgroup\else#1\fi}

\providecommand{\tabularnewline}{\\}

\@ifundefined{textcolor}{}
{%
 \definecolor{BLACK}{gray}{0}
 \definecolor{WHITE}{gray}{1}
 \definecolor{RED}{rgb}{1,0,0}
 \definecolor{GREEN}{rgb}{0,1,0}
 \definecolor{BLUE}{rgb}{0,0,1}
 \definecolor{CYAN}{cmyk}{1,0,0,0}
 \definecolor{MAGENTA}{cmyk}{0,1,0,0}
 \definecolor{YELLOW}{cmyk}{0,0,1,0}
}

\makeatother

\makeatother

\usepackage{babel}

\makeatother

\usepackage{babel}

\makeatother

\usepackage{babel}

\usepackage{babel}

\makeatother

\usepackage{babel}
\begin{document}

\title{Intrinsic switching field distribution of arrays of Ni$_{80}$Fe$_{20}$
nanowires probed by $\mathit{in\, situ}$ magnetic force microscopy}

\author{M. R. Tabasum$^{1}$, F. Zighem$^{1,2}$, J. De La Torre Medina$^{3}$,
L. Piraux$^{1}$ and B. Nysten$^{1}$}

\affiliation{$^{1}$ Institute of Condensed Matter and Nanosciences (IMCN), Université
catholique de Louvain, 1348 Louvain La Neuve, Belgium}

\affiliation{$^{3}$ Laboratoire des Sciences des Procédés et des Matériaux, CNRS-Université
Paris 13-Sorbonne Paris Cité, 93430 Villetaneuse, France}

\affiliation{$^{2}$ Facultad de Ciencias Físico Matemáticas, Universidad Michoacana
de San Nicolás de Hidalgo, Mexico}
\begin{abstract}
The progress of magnetization reversal of weakly packed ferromagnetic
Ni$_{80}$Fe$_{20}$ nanowire arrays of different diameters (40, 50,
70 and 100 nm) electrodeposited in polycarbonate membranes was studied
by magnetic force microscopy (MFM). For such a low packing density
of nanomagnets, the dipolar interactions between neighbouring wires
can be neglected. The intrinsic switching field distribution has been
extracted from in situ MFM images and its width was found to be considerably
smaller than for densely packed nanowire arrays.
\end{abstract}


\maketitle

\section{Introduction}

Arrays of magnetic nanostructures, such as of nanowires (NWs), are
extensively investigated due to their potential applications in magnetic
storage {[}1{]} and microwave devices {[}2{]}. Amongst the various
issues at stake for a comprehensive understanding of these arrays
is the influence of long range dipolar interactions {[}3-10{]} because
these interactions between NWs strongly influence the switching field
distributions (SFD) which play a significant role for information
storage. Particularly, the width of the SFD is important since smaller
values of this parameter leads to less recording errors and it is
also a measure of the quality of the recording media {[}11{]}. Consequently,
understanding and evaluation of the distinct influences of interactions
between NWs and intrinsic switching field distribution (SFD) in the
magnetization reversal process of such arrays is critical for the
development of magnetic recording media. The intrinsic SFD mostly
originates from non-uniformities of the geometrical parameters such
as the aspect ratio and shape of the NW tip that may affect the magnetization
reversal properties. The present work aims at the determination of
the magnetization reversal progress in low density two-dimensional
arrays of ferromagnetic NWs of different diameters (40, 50, 70 and
100 nm) grown by electrodeposition inside low porosity ($P<1\%$)
polycarbonate nanoporous membranes, by using in situ magnetic force
microscopy (MFM). In these arrays, the NWs are sufficiently isolated
from each other to neglect the dipolar interactions between them.
This helps to avoid the difficult corrections which are necessary
for dense NWs arrays {[}7-10{]}, leading to an easier analysis and
interpretation of the MFM experiments.

\section{Experimental procedure}

Arrays of Ni$_{80}$Fe$_{20}$ NWs have been fabricated by electrodeposition
in nanoporous polycarbonate (PC) templates of thickness 20$\mu$m.
The fabrication method of PC template, which has porosity less than
$1\%$, has been reported elsewhere {[}13{]}. Prior to the electrodeposition,
a gold layer was evaporated on one side of the membrane in order to
cover the pores and to use it as a cathode. In addition, for proper
adhesion of gold layer with the PC membrane a thin film of Cr ($\sim10$
nm) was deposited. The solution used to fabricate the Ni$_{80}$Fe$_{20}$
NWs contains: NiSO$_{4}$.6H$_{2}$O 131g/L, FeSO$_{4}$.6H$_{2}$O
5.56 g/L, H$_{3}$BO$_{3}$ 24.7g/L {[}14{]}. To ensure proper MFM
experiments; the Au and Cr layers were removed by chemical etching,
after electrodeposition, in order to obtain a smooth surface where
all the NWs tips at that side of the PC membrane are close to the
surface.

\begin{figure}
\includegraphics[bb=35bp 220bp 762bp 590bp,clip,width=8.5cm]{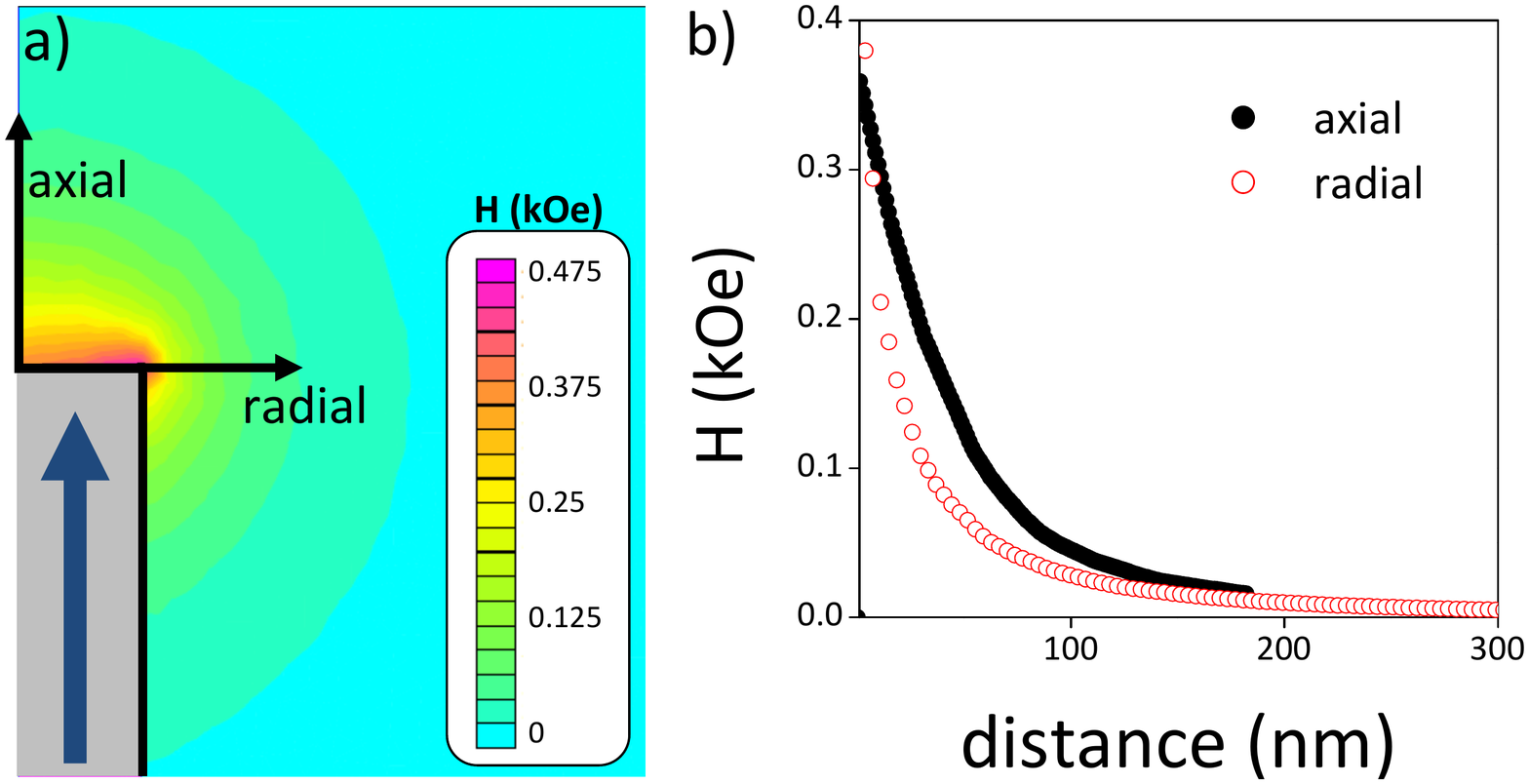} 

\caption{Dipolar fi{}eld emerging from a NW of diameter $D=100$ nm uniformly
magnetized $(4\pi M_{S}=10$ kG) calculated using \textsc{femm} {[}11{]}:
a) axial mapping of the field around the NW and$4\pi M_{S}=10$ kG)
calculated using \textsc{femm} {[}11{]}: a) axial mapping of the
field around the NW and b) magnitude of the field along the NW axis
and the radial axis}
\end{figure}

The length of the NWs is $L=4$ $\mu$m with a standard deviation
of $\pm0.1$ $\mu$m as determined by SEM. The pores diameters ($D$)
vary from 40 to 100nm with a standard deviation of the diameter $\sigma_{D}$
of $\pm3$ nm. In this range of $D$, the magnetic moment is expected
to be uniform inside each NW. Actually, the domain wall width $\delta_{wall}=\pi\sqrt{A/K}$
(where $A$ is the exchange stiffness and $K$ is the magnetocristalline
anisotropy) for Ni$_{80}$Fe$_{20}$ is few hundreds of nm. Consequently,
two remanent states are expected to be stable and correspond to a
uniform magnetization aligned along $+Oz$ and $\lyxmathsym{\textendash}Oz$
($+Oz$ being the NW\textquoteright{}s axis direction). Table I presents
the template porosities P and the mean interwire distances of the
first four neighbouring NWs. Figure 1 shows the dipolar stray fi{}eld
at the tip of a uniformly magnetized NW ($L=4$ $\mu$m, $D=10$0
nm, $4\pi M_{S}=10$ kOe) calculated using \textsc{femm} {[}11{]}.
It illustrates that the dipolar fi{}eld can be neglected in our arrays
since it is localized around the tip of the NWs: in a volume with
a typical size given by the radius of the NW and because the inter-wire
distances measured in our arrays is around 1000 nm.

\begin{figure}
\begin{centering}
\includegraphics[bb=0bp 200bp 500bp 750bp,clip,width=8.5cm]{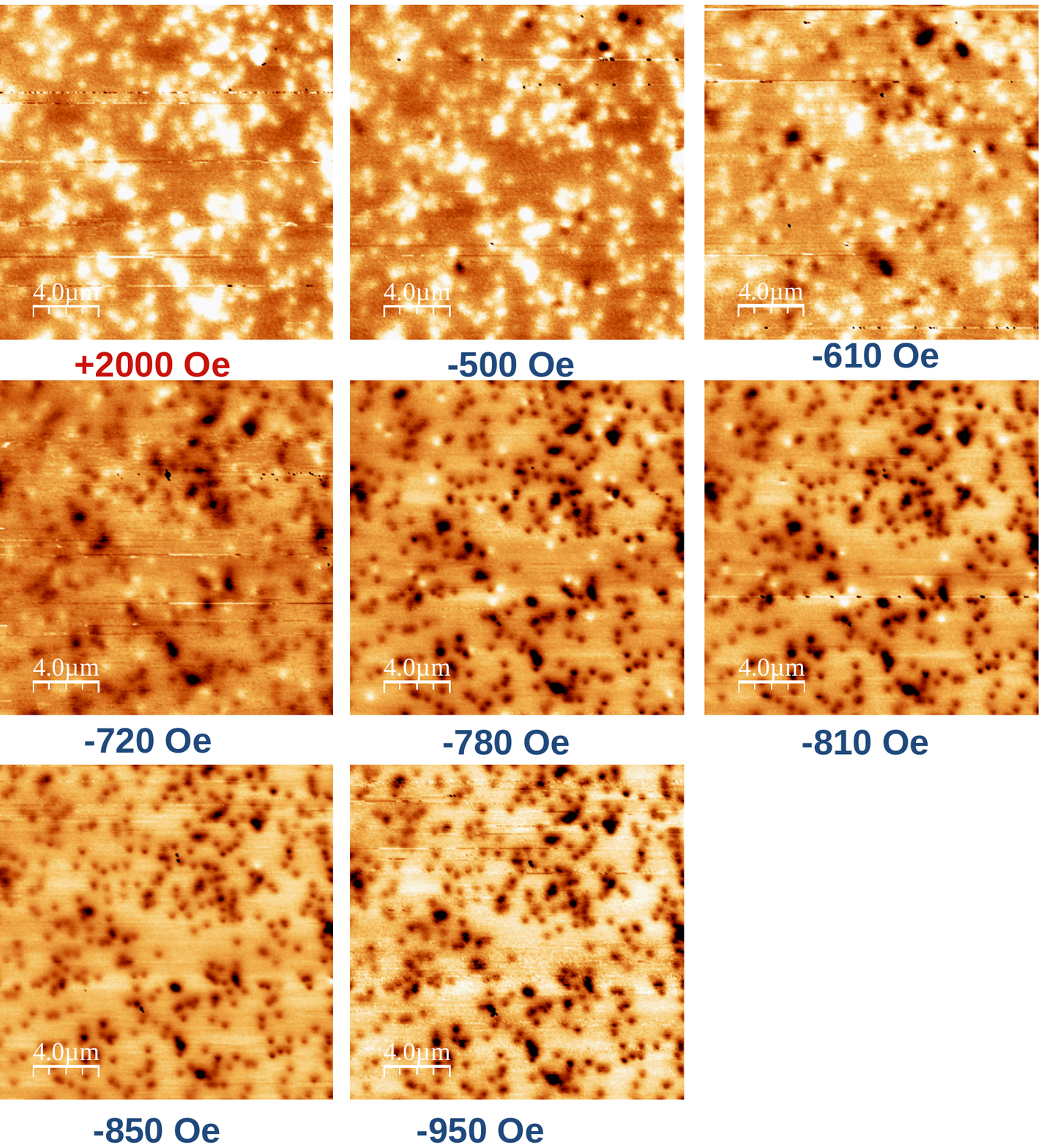} 
\par\end{centering}

\caption{Array of NWs with $D=50$ nm: series of MFM images (taken at zero
applied field: see text) indicating the progress of magnetization
reversal after saturation in a positive field ($H=+2000$ Oe) and
for different negative applied fields.}
\end{figure}

\begin{table}
\begin{tabular}{>{\centering}p{2cm}>{\centering}p{2cm}>{\centering}p{1cm}>{\centering}p{1cm}}
\multicolumn{4}{c}{}\tabularnewline
\hline 
\hline 
$D$ 

(nm) & Measured $P$ ($\%$) & \multicolumn{2}{>{\centering}p{2cm}}{Interwire distance (nm) }\tabularnewline
\hline 
\hline 
40 & 0.04 & \multicolumn{2}{>{\centering}p{2cm}}{$\sim$ 1200}\tabularnewline
50 & 0.15 & \multicolumn{2}{>{\centering}p{2cm}}{$\sim$ 1000}\tabularnewline
\multicolumn{1}{>{\centering}p{2cm}}{70} & 0.19 & \multicolumn{2}{>{\centering}p{2cm}}{$\sim$ 1000}\tabularnewline
100 & 0.25 & \multicolumn{2}{>{\centering}p{2cm}}{$\sim$ 1200}\tabularnewline
\hline 
\hline 
 &  &  & \tabularnewline
\end{tabular}

\caption{Templates porosities $P$ and the mean inter-wire distances.}
\end{table}

The magnetization reversal was investigated using a magnetic force
microscope from Agilent Technology. The experiments were done using
non-contact mode allowing us to obtain magnetic and topographic images
simultaneously. The extremity of the NWs could be imaged separately
while keeping them in the template, as well as their magnetic state:
up or down. An external magnetic field was applied using a custom-built
electromagnet with the ability to produce a controllable magnetic
field (parallel to the NWs axis) of up to 2 kOe. This made it possible
to not only saturate the sample but also to measure different DC demagnetization
(DCD) remanent states. We applied an in situ external magnetic field
parallel to the NWs before each measurement. The MFM images have been
obtained at zero applied fields (in different remanent states) after
switching off the external field. Indeed, as a consequence of the
weak dipolar interactions between NWs in low porosity membranes, measurements
done at a field $H_{0}$ and at zero field after applying a field
$H_{0}$ are same (this was checked experimentally). By counting the
number of NWs which reversed their magnetization direction after each
increment of applied field, we obtained the switching field distribution
(SFD) and MFM-magnetization curves.

\begin{figure}
\includegraphics[bb=25bp 265bp 555bp 595bp,clip,width=8.5cm]{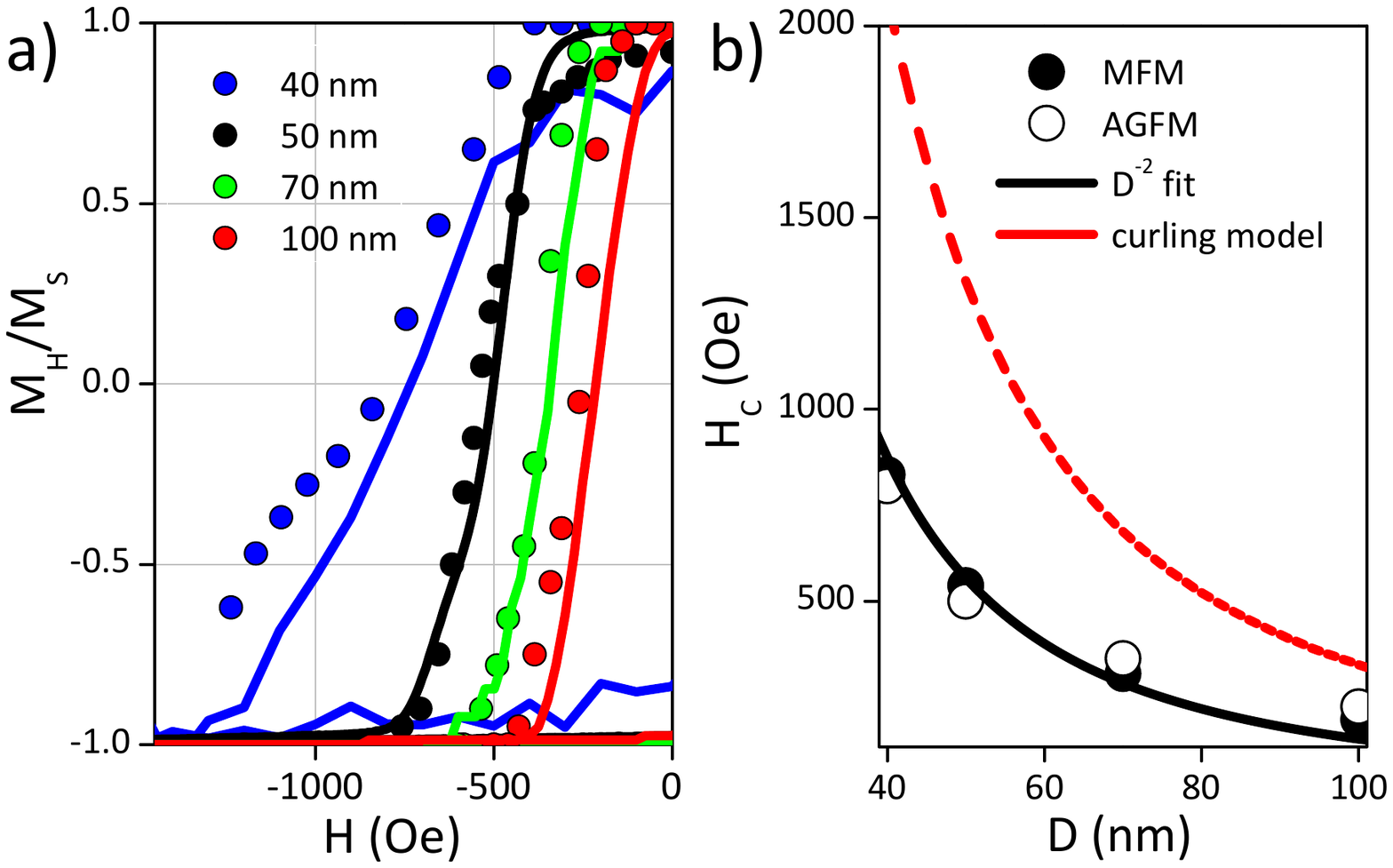} 

\caption{a) MFM-magnetization curves (solid dots) and bulk magnetization curves
(continuous lines) for the different arrays of NWs. b) Measured coercive
field $H_{C}$ as function of $D$, black continuous line corresponds
to a fit using a $D^{-2}$ law while the red dashed line is obtained
by using the analytical expression of $H_{C}$ in the \textquotedblleft{}curling
model\textquotedblright{} {[}15{]}. }
\end{figure}

\section{Experimental results and discussion}

Figure 2 presents a series of MFM images of the 50nm diameter NWs
array scanned at different remanent states. Before inserting the array
inside the microscope, it was saturated under a magnetic field of
$H=+$2 kOe along the NW axis ($+Oz$) while the magnetic tip of the
microscope was saturated in the opposite direction ($-Oz$). From
the first image (top left), which corresponds to the one just after
saturating the system, we observed that the NWs were still oriented
in the same direction (as seen from the white spots). Then, a series
of magnetic fields opposite to the sample's initial saturation field
was applied. The successive switching of the NWs started around $-500$
Oe till the application of $-950$ Oe where all the NWs had been switched
black. The MFM-magnetization curves were determined by representing
the percentage of switched NWs as a function of each increment of
applied field. Figure 3a shows the MFM-magnetization curves obtained
from the different arrays (solid dots). In addition, bulk magnetization
curves have been obtained using an alternating gradient force magnetometer
(AGFM) and compared to the MFM-magnetization curves (continuous lines
in Figure 3a). The results from MFM and AGFM are in good agreement.
The small deviations that appear between these two measurements can
be attributed to the fact that a limited number of NWs is probed using
MFM (local study) while the whole set of NWs is probed using magnetometry
measurements.

The variation of $H_{C}$ as a function of $D$ is represented in
Figure 3b; it increases with decreasing $D$ and ranges from $H_{C}=200$
Oe for $D=100$ nm to $H_{C}=800$ Oe for $D=40$ nm. This dependence
of $H_{C}$ with diameter fits well with a $D^{-2}$ law. It is well
know that the $H_{C}$ of an infinite NW in the coherent rotation
model is given by $H_{C}=2\pi M_{S}=5$ kOe and do not depends on
the diameter {[}15{]}. Nevertheless, our NWs are not infinite and
have diameters larger than $D_{coh}=7.3\ell_{ex}\sim$36 nm ( being
the exchange length) which corresponds to the diameter beyond which
non uniform reversal modes such as curling and buckling can occur
{[}15,16{]}. We have calculated the $D$-dependence of $H_{C}$ in
the curling model (see Figure 3b) for infinite Ni$_{80}$Fe$_{20}$
NWs by using the bulk magnetic parameter; namely saturation magnetization:
$4\pi M_{S}=10$ kG and exchange stiffness: $A=1\times10^{-6}$erg.cm$^{-1}$
{[}14, 15{]}. Although the measured HC are two times smaller than
the ones found in the model, we observe the same $D^{-2}$ dependence
which shows that the reversal mechanism in our NWs resembles the curling
model. 

The SFD is linked with the number of switched NWs at each increment
of the applied field and have been determined as a function of the
applied fields; it is reported in Figure 4a. The SFD width ($\delta_{SFD}$)
corresponds to the difference between the field of first reversed
NW and the last one and is reported in Figure 4b as a function of
$D$. It decreases with increasing $D$, from 800 Oe for $D=40$ nm
to 320 Oe for $D=100$ nm. These values are smaller than the ones
determined for densely packed NWs grown in alumina templates where
$\delta_{SFD}$ was more than 2 kOe (for $D$ around 40 nm) {[}7,
9, 10{]} and where strong dipolar interactions were present. It has
been shown that the dipolar interactions in arrays of NWs with a fixed
$D$ and increasing $P$ lead to a broadening of $\delta_{SFD}$ {[}4{]}.
Whereas, in our case, the packing density $P$ is relatively low ($<1
$) and decreases with decreasing $D$, this mean that the broadening
of $\delta_{SFD}$ with decreasing $D$ is mostly due to intrinsic
effects and not from the dipolar interactions. Let us assume that
$\delta_{SFD}$ is defined as $\delta_{SFD}=\delta_{SFD}^{D}+\delta_{SFD}^{0}$
where, $\delta_{SFD}^{D}$ is due to the size distribution of $D$
and $\delta_{SFD}^{0}$ may refer to; the inhomogeneities in the structure
and chemical composition of the NW arrays which contribute to the
broadening of the SFD. We propose to separately determine $\delta_{SFD}^{0}$
and $\delta_{SFD}^{D}$. For this purpose, we consider a Gaussian
size distribution of the NWs diameter with a standard deviation of$\pm5$
nm (see Figure 4c), as it agrees with the SEM measurements and we
use the $D^{-2}$ fit obtained from the experimental evolution of
$H_{C}$ with $D$ (Figure 3b). Figure 5a presents the calculated
by this way, it increases with decreasing $D$. The variation of $\delta_{SFD}^{D}$
and $H_{C}$ with $D$ are represented and fitted using $D^{-2}$
laws in Figure 5b. Note that $\delta_{SFD}^{0}$ is almost constant
with $D$ (blue squares in Figure 5b). This proves that the increase
of $\delta_{SFD}^{D}$ for smaller $D$ is essentially due to the
size distribution of the nanopores in the PC membrane while $\delta_{SFD}^{0}$
is due to other intrinsic effects.

\begin{figure}
\includegraphics[bb=28bp 290bp 443bp 585bp,clip,width=8.5cm]{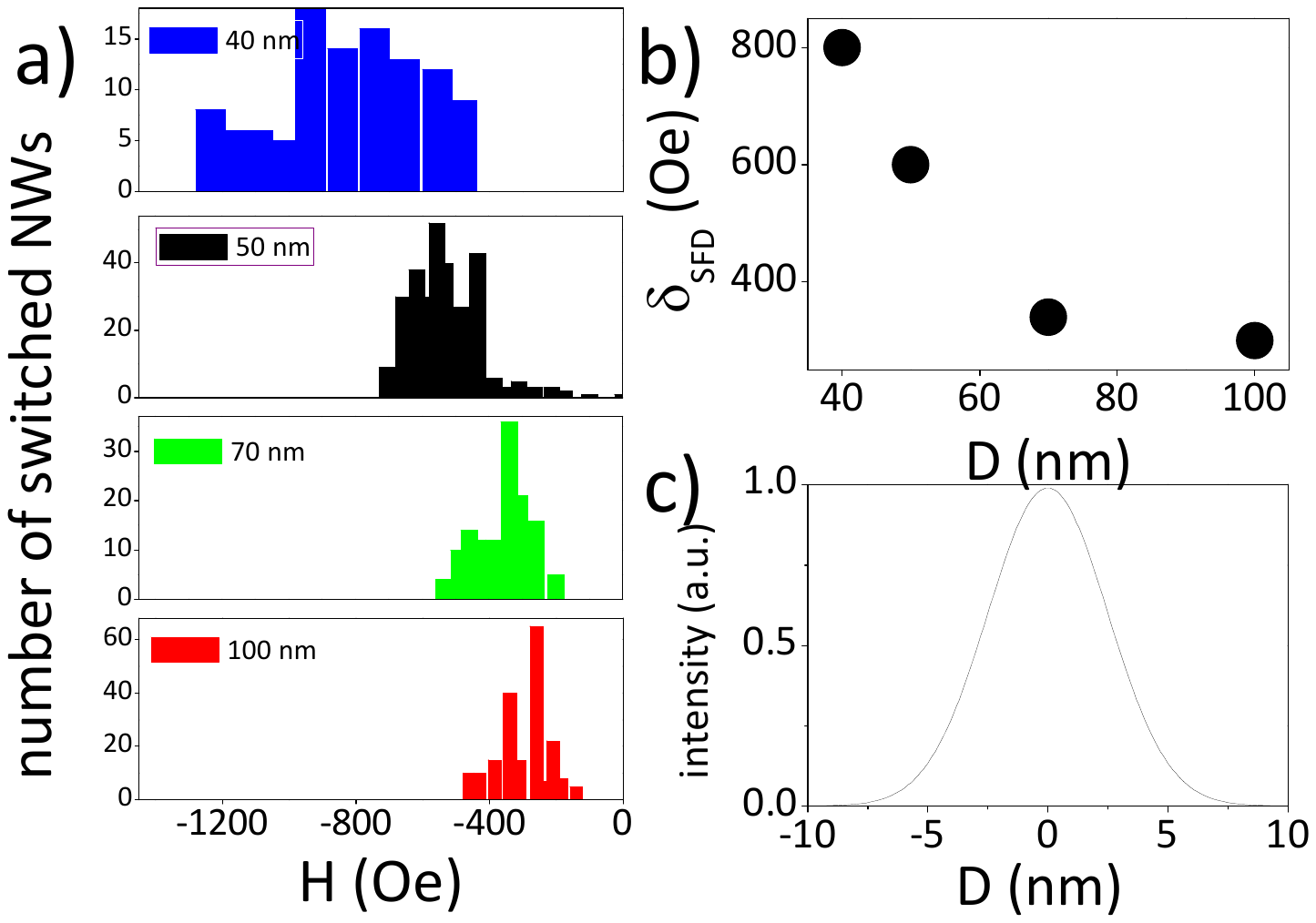} 

\caption{Switching field distributions of the different arrays of NiFe NWs.
b) Variation of the SFD width ($\delta_{SFD}$) as a function of $D$.
c) Gaussian distribution of the pores diameters.}
\end{figure}

\begin{figure}
\includegraphics[bb=30bp 205bp 585bp 575bp,clip,width=8.5cm]{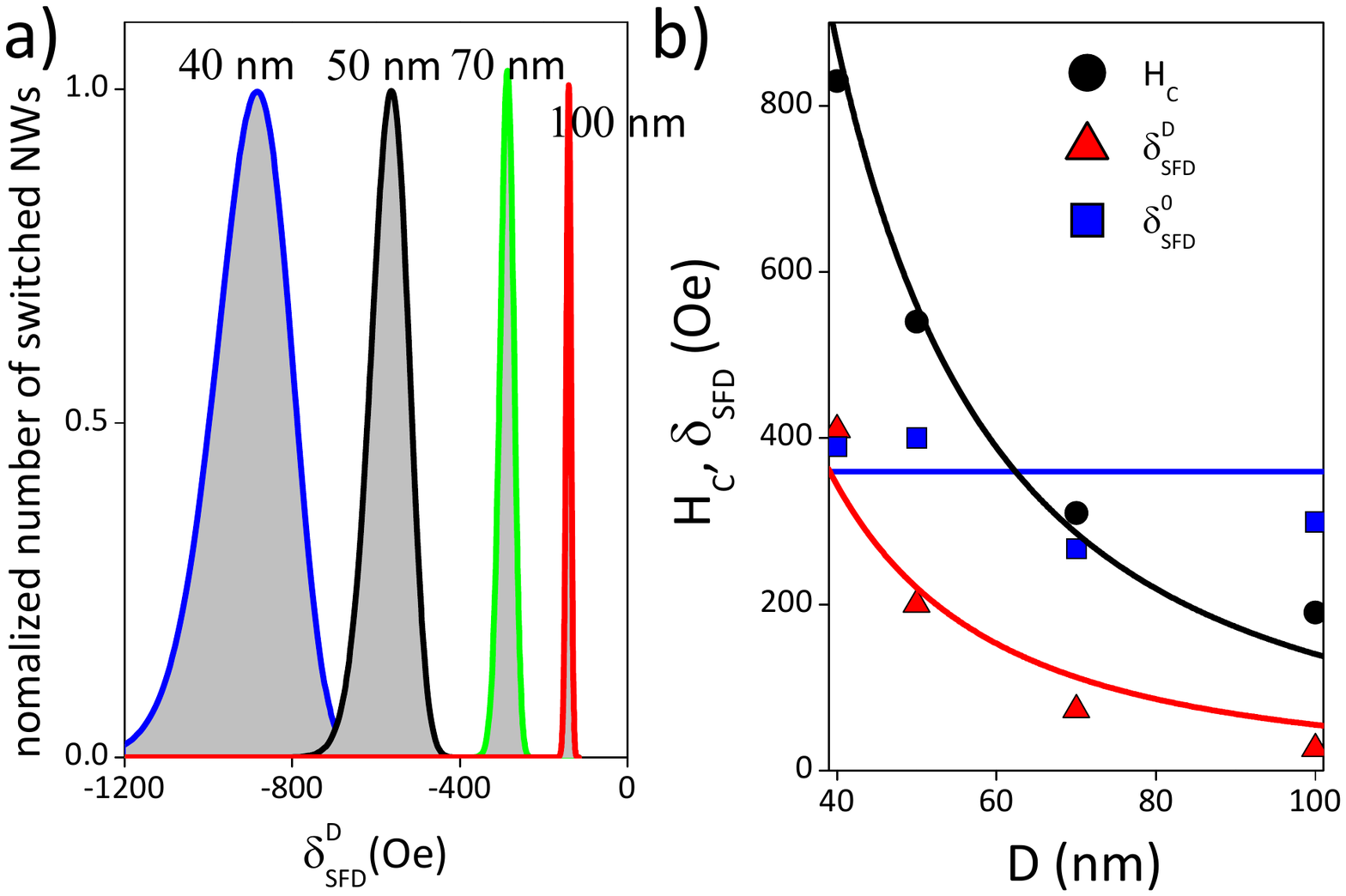} 

\caption{a) Calculated contribution of the size distribution of $D$ ($\delta_{SFD}^{D}$)
obtained from the Gaussian size distribution of the polycarbonate
nanopores and the $D^{-2}$ fit of $H_{C}$. b) Variation of $H_{C}$,
$\delta_{SFD}^{D}$and $\delta_{SFD}^{0}$ as a function of $D$.
The black and red continuous lines are fits using $D^{-2}$ law. }
\end{figure}

\section{Conclusion}

Arrays of Ni$_{80}$Fe$_{20}$ ferromagnetic nanowires embedded into
nanoporous polycarbonate membranes of low porosity have been studied
using in situ magnetic force microscopy. The influence of the diameter
of the nanowires on the magnetization reversal and thus the switching
field distribution (SFD) has been investigated. By counting number
of nanowires with magnetization up and down, the local magnetization
curves were obtained which agree well with bulk magnetization measurements.
We demonstrate narrow SFD width ($\delta_{SFD}$) regardless of the
nanowire's diameter ($D$) compared to the ones generally obtained
for densely packed nanowire arrays in alumina membranes. Moreover,
a decrease of $\delta{}_{SFD}$ with $D$ is observed and has been
qualitatively explained by considering a Gaussian distribution of
the nanopore diameter in polycarbonate membranes. The MFM magnetization
curves and the SFD of the samples show that such systems would be
suitable for data storage applications. However, the variation of
$\delta_{SFD}^{D}$ with $D$ could be a limitation if one wants to
use such systems with small diameter.

Arrays of Ni$_{80}$Fe$_{20}$ ferromagnetic NWs embedded into nanoporous
polycarbonate membranes of low porosity have been studied using in
situ magnetic force microscopy. The influence of the diameter of the
NWs on the magnetization reversal and thus the switching field distribution
(SFD) has been investigated. By counting number of NWs with magnetization
up and down, the local magnetization curves were obtained which agree
well with bulk magnetization measurements. We demonstrate narrow SFD
width ($\delta_{SFD}$) regardless of the nanowire's diameter (D)
compared to the ones generally obtained for densely packed ferromagnetic
NW arrays in alumina membranes. Moreover, a decrease of $\delta{}_{SFD}$
with $D$ increasing diameter is observed and has been qualitatively
explained by considering a Gaussian distribution of the nanopores
diameter. The variation of with $D$ could be a limitation if one
wants to use such systems with small diameter for data storage applications.
\begin{acknowledgments}
The authors thank E. Ferain for providing the PC templates. BN is
Senior Research Associate of the F.R.S.-FNRS of Belgium. M.R.Tabasum
is on leave from IME-RCET (University of Engineering and Technology
Lahore).\end{acknowledgments}


\begin{thebibliography}{References}
\bibitem{sellmyer2006} D.J. Sellmyer, Y. Xu, M. Yan, Y. Sui, J. Zhou
and R. Skomski J. Magn. Magn. Mater. \textbf{303,} 302 (2006)

\bibitem{aimad2005} A. Saib, M. Darques, L. Piraux, D. Vanhoenacker-Janvier
and I. Huynen IEEE Trans. Microw. Theo. Tech. \textbf{53} 2043 (2005)

\bibitem{encinas2001} A. Encinas Oropesa, M. Demand, L. Piraux, I.
Huynen, U. Ebels Phys. Rev. B \textbf{63}, 104415 (2001)

\bibitem{zighem2011} F. Zighem, T. Maurer, F. Ott and G. Chaboussant
J. Appl. Phys. \textbf{109}, 013910 (2011)

\bibitem{yelon} L. Clime, P. Ciureanu and A. Yelon, J. Magn. Magn.
Mater. \textbf{297}, 60 (2006)

\bibitem{nielsch}K. Nielsch, R. B. Wehrspohn, J. Barthel, J. Kirschner
and U. Gösele Appl. Phys. Lett. \textbf{79}, 1360 (2001)

\bibitem{medina_2010}J. De La Torre Medina, L. Piraux, J. M. Olais
Govea and A. Encinas Phys. Rev. B \textbf{81}, 144411 (2010)

\bibitem{Escrig} J. Escrig and D. Altbir, M. Jaafar, D. Navas, A.
Asenjo, and M. Vázquez, Phys. Rev. B 75, 184429 (2007)

\bibitem{asenjo_2006} A.Asenjo, M. Jaafar, D. Navas and M. Vazquez
J. Appl. Phys. \textbf{100}, 023909 (2006)

\bibitem{Sorop} T. G. Sorop, C. Untiedt, F. Luis, M. Kröll, M. Ra\c{s}a
and L. J. de Jongh Phys. Rev. B \textbf{67}, 014402 (2003)

\bibitem{FEMM} D. C. Meeker, Finite Element Method Magnetics (http://www.femm.info)

\bibitem{Lau_2011}J. W. Lau and J. M. Shaw, J. Phys. D: Appl. Phys.
\textbf{44}, 303001 (2011)

\bibitem{piraux97}R. Ferré, K. Ounadjela, J. M. George, L. Piraux
and S. Dubois Phys. Rev. B \textbf{56}, 14066 (1997)

\bibitem{encinas} A. Encinas, M. Demand, L. Vila and L. Piraux, Appl.
Phys. Lett. 81, 2032 (2002)

\bibitem{aharoni_1958}A. Aharoni and S. Shtrikman Phys. Rev. \textbf{109},
1522 (1958)

\bibitem{sellmyer_2000}R. Skomski, H. Zeng, M. Zheng, and D. J. Sellmyer
Phys. Rev. B \textbf{62}, 3900 (2000)

\bibitem{darques_2011} S. Da Col, M. Darques, O. Fruchart and L.
Cagnon Appl. Phys. Lett. \textbf{98}, 112501 (2011)\end{thebibliography}
\end{document}